\documentclass[a4paper]{jpconf}
\usepackage{graphicx}
\usepackage{dcolumn}
\usepackage{amssymb}
\usepackage{amsmath}
\usepackage{bm}
\newcommand{\ud}{\,\mathrm{d}}

\begin{document}

\title{Non-Collinear Ferromagnetic Luttinger Liquids}

\author{N.~Sedlmayr${}^1$, S.~Eggert${}^{1,2}$, J.~Sirker${}^{1,2}$}

\address{
${}^1$Fachbereich Physik and ${}^2$Optimas Research Centre, Technische Universit\"at Kaiserslautern,
Erwin-Schr\"odinger-Strasse Geb\"aude 46,
67663 Kaiserslautern, Germany}

\ead{sedlmayr@physik.uni-kl.de}

\begin{abstract}
The presence of electron-electron interactions in one dimension profoundly changes the properties of a system. The separation of charge and spin degrees of freedom is just one example. We consider what happens when a system consisting of a ferromagnetic region of non-collinearity, \emph{i.e.}~a domain wall, is coupled to interacting electrons in one-dimension (more specifically a Luttinger liquid). The ferromagnetism breaks spin charge separation and the presence of the domain wall introduces a spin dependent scatterer into the problem.
%Despite the absence of spin-charge separation we still have a system very different from a Fermi liquid.
The absence of spin charge separation and the effects of the electron correlations results in very different behaviour for the excitations in the system and for spin-transfer-torque effects in this model.
\end{abstract}

\section{Introduction}

The behaviour of the electronic and magnetic degrees of freedom in quasi one-dimensional wires and films has already received considerable interest and study, see for example the review of Marrows\cite{Marrows2005}. So far most works focus on how the transport properties of free electrons behave in a ferromagnetic wire with a domain wall, and how these spin polarized currents set the domain wall itself into motion. One obvious question to ask is whether interactions are important in such cases. A straightforward yes was answered by Dugaev et al.\cite{PhysRevB.65.224419}~who explicitly demonstrated that one must include at least mean field interaction corrections to correctly describe the charge build up around the domain wall. This of course should not be surprising but it is a timely reminder that correlation effects are often important even for nominally ``free'' electron systems. In this case failure to include these correlation effects results in an un-physical charge excess. In addition to the mean field interaction work of Dugaev et al.~there also exists in the literature some consideration of Hartree-Fock corrections\cite{PhysRevB.74.224429,PhysRevB.76.205107}. Here we will consider a more strongly correlated system: the Tomonaga-Luttinger liquid.

Experimentally\cite{Gambardella} the construction of chains of single magnetic atoms is already possible and one dimensional magnetic systems are experimentally available. Several obvious questions must be answered at this point. Traditionally it is thought that long range ferromagnetic order does not exist in one dimension. However, there are several things which mitigate this bald statement for these systems. The long range ferromagnetic order is in fact broken by the presence of domain walls, precisely the situation we are interested in. We also note that effects from the substrate mean that one can not really consider the chain as an isolated spin chain to be solved by a Heisenberg or Ising type model. Experimentally ferromagnetically ordered chains of atoms are indeed observed\cite{Gambardella,PhysRevB.56.2340,PhysRevLett.73.898,PhysRevB.57.R677}. Furthermore they exhibit non-collinear ferromagnetic order\cite{RevModPhys.81.1495}. One should of course bear in mind that the condition for defining the system as one-dimensional
may be different for the magnetic degrees of freedom and for the electronic degrees of freedom. It is only the latter that we require to be one-dimensional to be in the Luttinger regime.

These considerations lead us to our model of a ferromagnetic Tomonaga-Luttinger model\cite{Giamarchi,PTP.5.544,luttinger:1154,pereira04}. Unfortunately the chains of atoms do not easily display Luttinger physics due to the interference of the substrate\cite{0953-8984-13-22-301}, though they remain an intriguing possible future application of the model in a more complicated scenario. Nonetheless there is still a class of possible candidates left for our model: dilute magnetic semiconductors\cite{RevModPhys.78.809}. Systems where the magnetic and electronic degrees of freedom belong to different layers would also be a possible realization.

We will consider an ``s-d'' like model in which the bulk magnetization and the conduction electrons are treated as separate (though of course still interacting). As such we introduce two timescales into the problem, a fast electronic one and a slow magnetic one. This allows us to first answer the question of how the presence of the domain wall affects the Luttinger liquid, forgetting for the moment the effect the \emph{moving} domain wall will have on the conduction electrons.

The question of a non-collinear ferromagnetic Luttinger liquid has not yet been fully addressed\cite{pereira04}, we set up the model in section \ref{mode}. Firstly we would like to analyze the low energy effective field theories and their low temperature properties, the results of this are presented in section \ref{lowt}. Secondly we wish to understand the behaviour and nature of the appropriate excitations of the model. Finally we want to consider how this affects the magnetization dynamics of the domain wall. Due to the radically different nature of the excitations of a Luttinger, as opposed to Fermi, liquid we expect these dynamics to be very different. We offer some thoughts on this based on our preliminary results in section \ref{outlook} later in this article.

\section{The Model}\label{mode}

The magnetization of the wire is described by $\vec{M}(x)=M_s\cos[\Theta(x)]\hat{\mathbf{z}}+M_s\sin[\Theta(x)]\hat{\mathbf{y}}$ where
$M_s$ is the saturation magnetization and, using $\cos[\Theta(x)]=-\tanh[x/\lambda]$, we have a description of a domain wall of length $\lambda$ situated at $x=0$. This magnetization is coupled to the conduction electrons with a strength given by the exchange coupling $J$. Naturally we consider a screened, and hence short range, interaction $V(x-x')$. This allows us to start from the following standard ``s-d'' Hamiltonian\cite{Blundell}:
\begin{eqnarray}
\tilde{H}=\sum_\sigma\int\ud x\tilde{\psi}^\dagger_\sigma(x)\bigg[-\frac{1}{2m}\partial^2_{x}-\mu\bigg]\tilde{\psi}_\sigma(x)-J\sum_{\sigma\sigma'}\int\ud x\tilde{\psi}^\dagger_\sigma(x)\tilde{\psi}_{\sigma'}(x)\vec{\sigma}_{\sigma\sigma'}.\vec{M}(x)\nonumber\\
+\underbrace{\frac{1}{2}\sum_{\sigma\sigma'}\int\ud x\ud x'\tilde{\psi}^\dagger_\sigma(x)\tilde{\psi}^\dagger_{\sigma'}(x')V(x-x')\tilde{\psi}_{\sigma'}(x')\tilde{\psi}_\sigma(x)}_{=H_I}.\label{fullhamiltonian}
\end{eqnarray}
$\mu$ is the chemical potential.

In order to be able to linearize the system our first step must be to remove the spatially dependent, and in principle perhaps very large, magnetization. To this end we rotate the spin direction to get a collinear ferromagnet via the following gauge transformation\cite{PhysRevLett.78.3773,PhysRevB.16.4032}: $\mathbf{H}=\mathbf{U}^\dagger(x)\tilde{\mathbf{H}}\mathbf{U}(x)$, $\psi_{\sigma}(x)=U^\dagger_{\sigma\sigma'}(x)\tilde{\psi}_{\sigma'}(x)$, and $\mathbf{U}(x)=e^{\frac{i}{2}\theta(x)\bm{\sigma}^x}$. The interaction is left unaffected, the magnetization is locally rotated to a Zeeman term, and the kinetic energy operator introduces a new potential when it acts on the local rotation. Thus we have
\begin{eqnarray}
H&=&\sum_\sigma\int\ud x\psi^\dagger_\sigma(x)\bigg[-\frac{1}{2m}\partial^2_{x}-JM_s\sigma^z_{\sigma\sigma}-\mu\bigg]\psi_\sigma(x)+H_I+H_w\textrm{ and}\\
H_w&=&-\frac{1}{2m}\sum_{\sigma\sigma'}\int\ud x\psi^\dagger_\sigma(x)\big[i\theta'(x)\sigma^x_{\sigma\sigma'}\partial_x +\frac{i\theta''(x)}{2}\sigma^x_{\sigma\sigma'} -\frac{1}{4}[\theta'(x)]^2\delta_{\sigma\sigma'}\big]\psi_{\sigma'}(x).\label{hw}
\end{eqnarray}
This new potential can be approximated if we assume that the  Fermi wavelength $\lambda_F\ll\lambda$.  In this case only the first term of equation \eqref{hw} is relevant, the next two terms are of order $(\lambda_F/\lambda)^2$.

The next step is linearization via the ansatz $\psi_\sigma(x)=e^{ik_{F\sigma}x}\psi_{R\sigma}(x)+e^{-ik_{F\sigma}x}\psi_{L\sigma}(x)$, where $k_{F\sigma}=\sqrt{2m(\mu+JM_s\sigma^z_{\sigma\sigma})}$. The $R$ and $L$ indices denote right and left moving electrons respectively. Note that if the Zeeman term is large we must linearize around the spin split Fermi points. This leads to the physically relevant phenomena of the breakdown of spin-charge separation\cite{spincharge}.
Umklapp processes scattering two left movers into right movers and vice versa are always neglected here due to the non-commensurate nature of the Fermi wavevectors.

Finally we have our model to be bosonized\cite{Giamarchi}. Naturally the bosonic fields satisfy $[\phi_\sigma(x),\Pi_\sigma(x')]=i\delta(x-x')$ where $\Pi_\sigma(x)=\partial_x\theta_\sigma(x)$.
We first rotate to the spin and charge degrees of freedom. We can then diagonalize the quadratic part of the Hamiltonian with $\phi_{c/s}(x)=[\phi_{1}(x)\pm\phi_{2}(x)]/\sqrt{2}$ (and similar for the $\theta(x)$ fields) giving us for the quadratic part of the bosonic Hamiltonian
\begin{eqnarray}
H_0=\int\ud x\bigg[
[\partial_x\theta_c(x)]^2\frac{v_cK_c}{2}
+[\partial_x\theta_s(x)]^2\frac{v_sK_s}{2}
+[\partial_x\theta_c(x)\partial_x\theta_s(x)] v_a\nonumber\\
+[\partial_x\phi_c(x)]^2\frac{v_c}{2K_c}
+[\partial_x\phi_s(x)]^2\frac{v_s}{2K_s}
+[\partial_x\phi_c(x)\partial_x\phi_s(x)] v_b
\bigg].
\end{eqnarray}
$K_s$ and $K_c$ are the spin and charge Luttinger parameters. $v_a$ and $v_b$ describe the coupling between the spin and charge sectors. These parameters are functions of the interaction strengths and Fermi velocities.

The diagonalization of the quadratic terms introduces new velocities, $u_i$,
for $H_0$ and a set of $\{T_i^{\theta,\phi}\}$ parameters. These are all known in terms of the previously mentioned Luttinger parameters.
The final Hamiltonian is $H=H_0+H_1+H^f_w+H^b_w$ where
\begin{eqnarray}
H_0&=&\int\ud x\sum_{i=1,2}\frac{u_i}{2}[(\partial_x\tilde{\phi}_i(x))^2+(\tilde{\Pi}_i(x))^2]\\
H^f_w&=&\frac{1}{2\pi\alpha}\int\ud x [  v_{F\uparrow}+  v_{F\downarrow}]\frac{\textrm{sech}[x/\lambda]}{\lambda} \sin[\sqrt{2\pi}(T^\theta_1\tilde{\theta}_1(x)+T^\theta_2\tilde{\theta}_2(x))]\nonumber\\&& \sin[x(k_{F\uparrow}-k_{F\downarrow})+\sqrt{2\pi}(T^\phi_1\tilde{\phi}_1(x)+T^\phi_2\tilde{\phi}_2(x)]\\
%\sin[x(k_{F\uparrow}-k_{F\downarrow})-\sqrt{2\pi}(T^\phi_1\tilde{\phi}_1(x)+T^\phi_2\tilde{\phi}_2(x)]\\
H^b_w&=&\frac{1}{2\pi\alpha}\int\ud x[  v_{F\uparrow}-  v_{F\downarrow}]\frac{\textrm{sech}[x/\lambda]}{\lambda} \sin[\sqrt{2\pi}(T^\theta_1\tilde{\theta}_1(x)+T^\theta_2\tilde{\theta}_2(x))]\nonumber\\&& \sin[x(k_{F\uparrow}+k_{F\downarrow})-\sqrt{2\pi}(\tilde{T}^\phi_1\tilde{\phi}_1(x)+\tilde{T}^\phi_2\tilde{\phi}_2(x))]\\
H_1&=&\frac{2}{(2\pi\alpha)^2}\int\ud x\cos[2\sqrt{2\pi}(T^\phi_1\tilde{\phi}_1(x)+T^\phi_2\tilde{\phi}_2(x))]\nonumber\\&&\big[[g_{1\perp\uparrow}+g_{1\perp\downarrow}]\cos[2x(k_{F\uparrow}-k_{F\downarrow})] -i[g_{1\perp\uparrow}-g_{1\perp\downarrow}]\sin[2x(k_{F\uparrow}-k_{F\downarrow})]\big].
\end{eqnarray}
We have both forward and backward scattering terms, $H^f_w$ and $H^b_w$ respectively. We also find an oscillating sine-Gordon interaction term $H_1$, describing ``slow'' Umklapp processes which are not averaged out, with some strength $g_{1\perp\sigma}$. These all couple our otherwise diagonal bosonic degrees of freedom. The appropriate excitations of such an $SU(2)$ asymmetric model have no obvious physical interpretation.
This summarizes the effective bosonic field theory which is the foundation of our mathematical analysis of the problem.

\section{Low Energy Physics}\label{lowt}

We start by writing a functional integral partition function\cite{NegeleOrland}:
\begin{equation}
\mathcal{Z}=\int D\tilde{\phi} D\tilde{\Pi} e^{-\int_0^{ \beta} \ud \tau\big[\int\ud x[-i\tilde{\Pi}_i(x)\partial_\tau\tilde{\phi}_i(x)]+H[\tilde{\Pi}(x),\tilde{\phi}(x)]\big]}
\end{equation}
with periodic boundary conditions in imaginary time $\tau$. Following the standard procedure we split the fields into fast, $\phi^>$, and slow, $\phi^<$, fields. Our fast fields are defined for $\Lambda'<|k|,|\omega|/u_i<\Lambda$, and the slow for $|k|,|\omega|/u_i<\Lambda'$. Expanding the exponent in terms of $H_1$ and $H_w$
and performing the averaging over the fast modes we then re-exponentiate the expression to find the appropriate scaling equations.
Parameterizing $\Lambda^({}'{}^)=\Lambda_0e^{-l(-dl)}$ we find for the first order $H_1$ term
\begin{eqnarray}
\frac{dg_{1\perp\sigma}}{dl}=
%g_{1\perp\sigma}\big[2-4[(T^\phi_1)^2+(T^\phi_2)^2]\big]\equiv
g_{1\perp\sigma}\gamma^1.
\end{eqnarray}
$\gamma^1$, and also the $\gamma^{f,b}$ used below, are known functions of the rotation terms $\{T_i^{\theta,\phi}\}$. This term is an irrelevant perturbation for any situation we wish to look at. In the limit of weak magnetization we can simplify the expressions to find $\gamma^1\to2-2K_s$.
In this limit $K_s>1$ and it becomes clear that the term is irrelevant.

The second order equation for our model is more difficult than for the sine-Gordon model. A diagonal equation in the $\tilde{\phi}_i$'s is not recovered and to perform any further analysis we would have to rediagonalize the problem and then renormalize the model once again. This is perhaps not totally unexpected as the scattering term we are dealing with explicitly couples these terms. We leave the more involved second order renormalization group analysis to a future work and stick here to the first order equations.

The same analysis is performed on the scattering terms, we define the scattering coupling constants $g^{f/b}(0)\equiv v_{F\uparrow}\pm v_{F\downarrow}$. For back scattering we find:
\begin{eqnarray}
\frac{dg^b}{dl}=
%g^b\big[2-[(\tilde{T}^\phi_1)^2+(\tilde{T}^\phi_2)^2+(T^\theta_1)^2+(T^\theta_2)^2]\big]\equiv
g^b\gamma^b.
\end{eqnarray}
This term is a relevant perturbation. In the limit of weak magnetization this simplifies to $\gamma^b\to2-[(K_s)^{-1}+K_c]/2$.
Similarly the forward scattering equation stands as
\begin{eqnarray}
\frac{dg^f}{dl}=
%g^f\big[2-[(T^\phi_1)^2+(T^\phi_2)^2+(T^\theta_1)^2+(T^\theta_2)^2]\big]\equiv
g^f\gamma^f.
\end{eqnarray}
This is also a relevant term. In the limit of weak magnetization this becomes $\gamma^f\to2-[(K_s)^{-1}+K_s]/2$.

In contrast to the scaling equations of Pereira and Miranda\cite{pereira04,pereirathesis} we find that we can not neglect forward scattering. They were interested in the limit of very sharp walls and the small magnetization limit. The model they solve leads to the consideration of a magnetic impurity in an otherwise collinear ferromagnetic system. In such a case it is perhaps not surprising that this model leads to the magnetic analog of the Kane-Fisher problem\cite{PhysRevB.46.10866,PhysRevB.46.15233}. Their final bosonic Hamiltonian is appropriate only for very sharp walls where the domain wall profile will be different to the longer walls which we wish to consider, which leads to the different scaling equations and hence a different effective low temperature model. It should also be pointed out that depending on the length of the domain wall forward scattering can become the \emph{dominant} form of scattering. Of course in the physical model ``forward scattering'' refers to an electron which passes through the wall \emph{without} changing its spin.

We have three natural length scales present in the problem: $\lambda_+=2\pi(k_{F\uparrow}+k_{F\downarrow})^{-1}$, $\lambda_-=2\pi(k_{F\uparrow}-k_{F\downarrow})^{-1}$, and $\lambda$. In the limit of weak magnetization $\lambda_-\to\infty$ and $2\lambda_+\to\lambda_F$ and the limit of large magnetization, when one spin channel becomes frozen out gives $\lambda_+\approx\lambda_-\approx\lambda_F$. We are generally interested in long domain walls when $\lambda\ge\lambda_F$ and we are in the physical regime where both forward and backward scattering are important. As one might expect in the large magnetization limit the scattering terms can be neglected as only one spin channel is available. For very sharp delta function like walls one finds only back scattering relevant\cite{pereira04}, a case we do not consider further. For long domain walls we are back in the adiabatic limit where scattering can be completely neglected. Interestingly there is a regime in between, $\lambda_+<\lambda<\lambda_-$, where forward scattering is the dominant mode of scattering. This regime can also be reached by tuning the magnetization to be suitably weak.

\section{Outlook}\label{outlook}

We are principally interested in the spin density and the spin and electronic currents. The spin density and current will also necessary for understanding the magnetization dynamics of the domain wall. The spin density\cite{PhysRevLett.75.934,PhysRevB.62.4370} is defined as $\vec{S}(x)={\bm\psi}^\dagger(x) {\bm U}^\dagger\vec{{\bm\sigma}}{\bm U}{\bm\psi}(x)$. It can be calculated perturbatively in the domain wall profile using the bosonic model derived above. Preliminary results show an important development. The corrections to the zeroth order density caused by scattering are concentrated around the edges of the domain wall. In a normal quasi one-dimensional wire these corrections are found around the centre of the domain wall\cite{PhysRevB.65.224419} and contribute to a spin torque which causes precessional motion of the domain wall, in analogy with a moment in a magnetic field. In contrast we expect the domain wall to be distorted rather than set into precessional motion by these terms in the Luttinger liquid. The temperature dependence of the spin density corrections can be extracted from these corrections for the small temperature limit. One finds power law contributions in the forward and backward scattering exponents: $T^{-\gamma^{f,b}}$. These become sharper for larger values of $JM$.

The magnetization dynamics are described by the Landau-Lifschitz-Gilbert equation, or some suitable generalization thereof\cite{LL9,Gilbert2004,PhysRevLett.93.127204}.
There are two different aspects to this. One is the straightforward point that the dynamics, \emph{over the length of the wall}, will be affected by the different spin density of the Luttinger liquid compared to the Fermi liquid or non-interacting case. The second, more interesting point, is whether the derivation of the non-adiabatic terms in the LLG equation are valid for a Luttinger liquid.

Following Zhang and Li\cite{PhysRevLett.93.127204} one can derive contributions to the magnetization dynamics which allow for the fact that the electrons do not instantaneously follow the magnetization profile. One first writes a continuity equation for the spins, assuming part of it to be always parallel to the bulk magnetization and allowing a small deviation from this. In order to derive the current dependent (so called $\beta$-) terms, those which drive the domain wall along the wire, one assumes that $j^{x,y,z}_s(x,t)=-\mu_BPj^{x,y,z}_e(x,t)M^{x,y,z}(x,t)/eM_s$. $\vec{j}_e(x,t)$ is the charge current and $P$ is the magnitude of the polarization, whilst $\vec{j}_s(x,t)$ is the spin current. A quite reasonable assumption in a Fermi liquid, this of course starts to look more dubious in the case of a Luttinger liquid. When $SU(2)$ symmetry is present, not our case one is reminded, spin and charge are of course uncorrelated and possess different velocities. Thus this assumption would completely fail. For us the situation is not so simple, nonetheless what is obvious from the model is that spin and charge are not fully correlated. One is forced to work with the spin current and not the electric current and, as we have already seen, the spin degrees of freedom can behave rather differently for this model.

Other than the magnetic behaviour we also wish to analyze the transport through the wire.

\section{Conclusion}

We have considered a non-collinear ferromagnetic Luttinger liquid. We assessed its low energy properties and the relevant lengthscales, presenting the results of renormalization group calculations. It is possible to tune the relevant types of scattering operators by considering different domain wall lengths and magnetization magnitudes. Here `magnetization' refers to both its magnitude and the magnitude of the exchange coupling. The regime of most interest to us which we go on to assess is the intermediate regime where we have both forward and backward scattering terms and an appreciable magnetization. We discuss some preliminary results for the spin density correlation function and the effect that this will have on the subsequent domain wall dynamics. We propose that in our Luttinger liquid system the magnetization dynamics will be altered by the very different nature of the Luttinger liquid excitations.

\subsection*{Acknowledgments}
The authors wish to thank J.~Berakdar and R.~Pereira for useful and stimulating discussions.

\section*{References}
\bibliography{jemsreferences}

\end{document}